\documentclass[amsmath, amssymb,aps, pre,twocolumn,showpacs]{revtex4-1}

\usepackage{amsmath}
\usepackage{url}
\usepackage{graphicx}
\usepackage{epstopdf}
\usepackage{color}
\usepackage{hyperref}
\usepackage{amsopn}
\usepackage{amssymb}
\usepackage{subfigure}
\newcommand{\mybinom}[2]{\Bigl(\begin{array}{@{}c@{}}#1\\#2\end{array}\Bigr)}
\DeclareMathOperator*{\MAX}{MAX}
\DeclareMathOperator*{\MIN}{MIN}
\DeclareMathOperator*{\SOR}{\bigvee}

\DeclareGraphicsExtensions{.pdf,.png}

\begin{document}

\title[Risk perception in multiplex networks]{Epidemic spreading and risk perception in multiplex networks: a self-organized percolation method}
\author{Emanuele Massaro}
\affiliation{Risk and Decision Science Team,
US Army Engineer Research and Development Center,
696 Virginia Rd., COncord, MA 01742, and Department of Civil and Environmental Engineering,
Carnegie Mellon University, 5000 Forbes Ave, Pittsburgh, PA 15213
}
\email{emassaro@andrew.cmu.edu}

\author{Franco Bagnoli}
\affiliation{Dipartimento di Fisica ed Astronomia and CSDC,  Universit\`a degli Studi di Firenze, via G. Sansone 1, 50019 Sesto Fiorentino, Italy. Also INFN, Sez. di Firenze.}
\email{franco.bagnoli@unifi.it}

\begin{abstract}
In this paper we study the interplay between epidemic spreading and risk perception on multiplex networks. The basic idea is that the effective infection probability is affected by the perception of the risk of being infected, which we assume to be related to the fraction of infected neighbours, as introduced by Bagnoli \textit{et al.}, PRE \textbf{76}:061904 (2007). We re-derive  previous results using a self-organized method, that automatically gives the percolation threshold in just one simulation. We then extend the model to multiplex networks considering that  people get infected by contacts in real life but often gather information  from an information networks, that may be quite different from the real ones. The similarity between the real and information networks determine the possibility of stopping the infection for a sufficiently high precaution level: if the networks are too different there is no mean of avoiding the epidemics. 
\end{abstract}


\maketitle
\section{Introduction}
Recently, the Health magazine reported ``Although H1N1 influenza killed more than 4,000 people in the United States in 2009-2010, this outbreak was relatively mild compared to some flu pandemics''~\cite{storia0}.

Indeed, the  twentieth century was characterized by a series of more serious events. During the 1918-19 the world assisted to the so-called \emph{Spanish Flu}. Starting from three different places: Brest (France), Boston (Massachusetts) and Freetown (Sierra Leone), the disease spread worldwide, killing 25 million people in 6 months (about 17 million in India, 500,000 in the United States and 200,000 in the United Kingdom).

In 1957, another pandemic originated in China and spread rapidly in Southeast Asia, taking hence the name of \emph{Asian}. The virus responsible was identified in the subtype H2N2, new to humans, resulting from a previous human H1N1 virus that was remixed with a duck virus from which it received the genes encoding the H2 and N2. This pandemic took eight months to travel worldwide and caused one to two million victims. 

The 1968 pandemic was the mildest of the twentieth century and started once again in China. From there it spread to Hong Kong, where more than half a million people fell ill, and in the same year reached the United States and the rest of the world.

Given these facts (and the whole records of pandemics in history~\cite{pandemic}), it is not surprising that the public health organizations are concerned about the appearance of a new deadly pandemics.    

However, in recent decades there have been many cases of false or exaggerated information about epidemics.  One example if the \emph{Swine flu} of 1976, or the  \emph{Avian flu} in 1997 where a United Nations health official warned that the virus could kill up to 150 million worldwide~\cite{storia0} or the more recent 2009 \emph{H1N1 flu}, during whose outbreak the U.K. Department of Health warned about 65000 possible deaths as reported by the Daily Mail in 2010~\cite{storia}. Fortunately these fears did not  realize. 

These catastrophic  scenarios  and  the extent of their impacts on the economic and social contexts induce  a reflection on the method used to forecast the evolution of a disease in real world. It is well known that in deeply connected networks (and in particular in scale-free ones without strong compartmentalization) the epidemic threshold of standard epidemic modeling is vanishing~\cite{epidemic1,epidemic2,epidemic3,epidemic4,epidemic5}. Indeed, the \emph{lazzarettos}~\cite{wikipedia} experience first in  Venice and then in many ports and cities was so successful in the absence of effective treatments because it was able to break the contact network.  Indeed, the pest was last observed in Venice in 1630, whereas in southeastern Europe, it was present until the 19th century~\cite{lazzaretti}. 

However, the last deadly pandemics of pest in Europe happened in 1820~\cite{pandemic}, and worldwide  in Vietnam in the 60's; the last pandemic influenza, Hong Kong flu, in 1968-1969.  In other words, the last rapid deadly pandemics happened well before the appearance of  highly-connected human networks.

Clearly,  the public health systems put a lot of efforts in trying to make people aware of the dangers connected to hygiene, dangerous sexual habits and so on. 
Indeed, the current worldwide diffusion of large-scale diseases (HIV, seasonal influenza, cold, papilloma virus, herpes virus, viral hepatitis among others) is deeply related to their silent (and slow) progression or to the assumption (possibly erroneous) or their harmlessness.   
However, it is well known that the direct experience (contact with actual ill people) is much more influential than public exhortations. 

Therefore, in order to accurately modeling the spreading of a disease in human societies, we need to take into account the perception of its impacts and the consequent precautions that people take when they become aware of an epidemic. These precautions may consist in changes in personal habits, vaccination or modifications of the contact network. 

We are here interested in diseases for which no vaccination is possible, so that the only way of avoiding a pandemic is by means of an appropriate level of precautions, in order to lower the infection rate below the epidemic threshold. We also assume that there is no acquired immunity from the disease and that its consequences are sufficiently mild not to induce a radical change in social contacts. 

In a previous work~\cite{riskperception}, some of us investigated the influence of the risk perception in epidemic spreading. We assumed that the knowledge about the diffusion of the disease among neighbors (without knowing who is actually infected) effectively lowers the probability of transmission (the effective  infectiousness). We studied the worst case of an infection on a scale-free network with exponent $\gamma =2$ and we showed that in this case no degree of prevention is able to stop the infection and one has to take additional precaution for hubs (such as public officers and physicians). 

We extend here the investigation to different network structures, on order to obtain a complete reference frame. For regular, random, Watts-Strogatz small-world and non-assortative scale-free networks with exponent $\gamma >3$ there is always a finite level of precaution parameter for which the epidemics go extinct. For scale-free networks with $\gamma < 3$ the precaution level depends on the cutoff of the power-law, which at least depends on the finite number of the network. 

We consider then an important factor of modern society: the fact that most of information comes no more  from physical contacts nor from broadcasting media, but rather from  the ``virtual'' social contact networks~\cite{virtual_inf, virtual_inf1, virtual_inf2}. A recent study, \emph{State of the news media} for the United States~\cite{article}, highlights this phenomena. It shows the extent of the influence of social networks, for what concerns subscribers who can read news published by newspapers. The $9\%$ of the population claims to inquire ``very often'' through Facebook and Twitter and  seven out of ten members are addressed to articles (from newspapers and other sources) by friends and family members.  

We are therefore confronted with news coming mainly from an information network. On the other hand, the real network of contacts is the environment where actual infections occur. We extend  our model to the case in which the source of information (mixed real and virtual contacts) does not coincide with the actual source of infection (the real contacts). 

This system is well represented as a multiplex network~\cite{multiplex1, multiplex2, multiplex3, multiplex4, multiplex5}, \textit{i.e.}, a graph composed by several  layers in which the same set of $N$ nodes can be connected to each other by means of links belonging to different layers, which represents a specific case of interdependent network~\cite{interdependent1, interdependent2}. 
Recently,  Granell et al.~\cite{multiplex9} have pointed out the attention to an interesting scenario where the multiplex corresponds to a two-layers network, one where the dynamics of the awareness about the disease (the information dynamics) evolves and another where the epidemic process spreads. They have also investigated the effect of \emph{mass media} when all the agents are aware of the infection~\cite{multiplexnuovo} which is the best case for stopping the epidemic. However, here we are interested in the studying a case of \emph{neglected diseases} in which the probability to be aware of the disease is very low (and not advertised by mass-media).

The first layer represents the information network where people become aware of the epidemic thanks to news coming from virtual and real contacts in various proportions. The second layer represents the real contact network where the epidemic spreading takes place. 

In this paper we want to model the effect of the virtual information for simulating the awareness of the agents in the real-world network contacts. We study how the percolation threshold of a  susceptible-infected-susceptible (SIS) dynamics depends on the  perception of the risk (that affects the infectivity probability) when this information comes from the same contact network of  the disease or from a different network. In other words, we study the interplay between risk perception and disease spreading on multiplex networks. 

We are interested in the epidemic threshold, which is a  quantity that it is not easy to obtain automatically (for different values of the parameters)  using numerical simulations. We extend  a self-organized formulation of percolation phenomena~\cite{bagnoli_rech} that allows to obtain this threshold in just one simulation (for a sufficiently large system). 
  \begin{figure}[t!]
\centering
{\includegraphics[width=0.95\columnwidth]{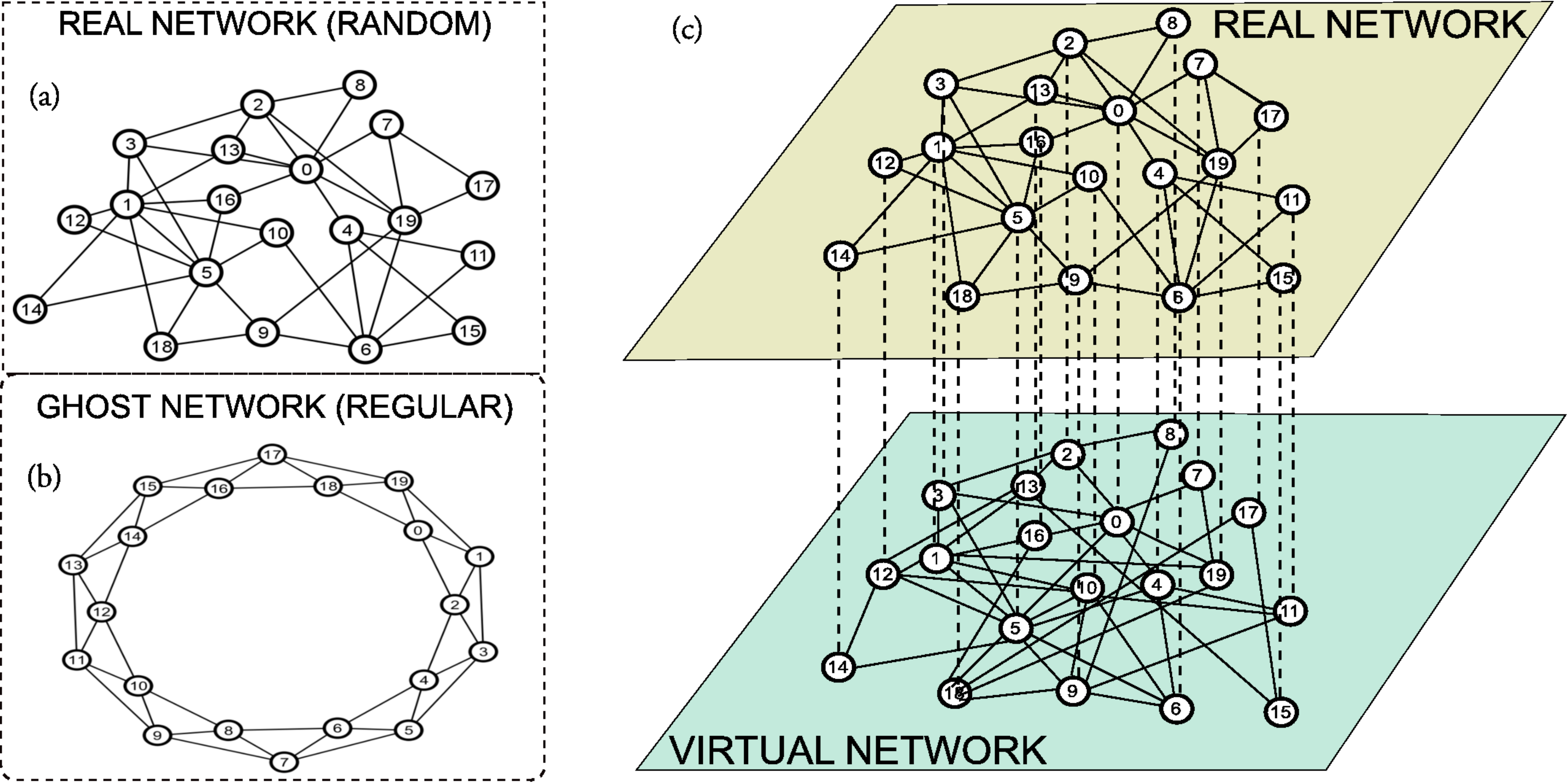}}
\caption{\label{fig:multiplex} Example of multiplexes generated with our method where (a) $R_{net}$ and (b) $G_{net}$ are both networks with $N=20$ nodes and $\langle k\rangle=4$. (c) Multiplex networks where the \emph{information} network ($V_{net}$) is given by the mixing of (a) and (b) with $p_r=0.5$.}
\end{figure}

\section{The network model}
\label{netsection}
In this section we show our method for generating multiplex networks. First of all we describe the  mechanisms for generating \emph{regular}, \emph{random} and \emph{scale-free}  networks. 

Let us denote by $a_{ij}=0,1$ the adjacency  matrix of the network, $a_{ij}=1$ if there is a link from $j$ to $i$ and zero otherwise. We shall denote by $k_i = \sum_j a_{ij}$ the connectivity of site $i$ and by $j^{(i)}_1, j^{(i)}_2, \dots, j^{(i)}_{k_i}$ that of its neighbours ($a_{i, j^{(i)}_n}= 1$). We shall consider only symmetric networks. We generate networks with $N$ nodes and $2mN$ links, so that the average connectivity of each node is $\langle k \rangle = 2m$.   
\begin{itemize}
\item Regular 1D: Nodes are arranged on a ring (periodic boundary condition). Any given node establishes a link with the  $m$ closest nodes at its right. For instance for $m=2$, node 1 establishes a link with nodes 2 and 3, node 2 with nodes 3 and 4, and so on until node $N-1$ establishes links with nodes $N$ and $1$ and node $N$ with nodes $1$ and $2$.
\item Random: Any node establishes $m$  links with randomly chosen nodes, avoiding self-loops and multiple links. 
The probability distribution of random networks is Possonian, $P(k) = \frac{z^ke^{-z}}{k!}$, where $z=\langle k \rangle$.
\item Scale-free: we use a configurational model fixing also a cutoff $K$. First, at each node $i$ out of $N$ is assigned a connectivity $k_i$ draft from a power-law distribution $P(k) = A k^{-\gamma}$, $m\le k \le K$, with $A=(\gamma-1)/(m^{1-\gamma}-K^{1-\gamma})$. Then links are connected at random avoiding  self-loops and multiple links, and finally the total number of link is pruned in order to adjust the total number of links. This mechanism allow us to generate scale-free networks with a given exponent $\gamma$. 
\end{itemize}

We are interested in multiplex networks composed by two sub-networks that we denote \emph{real}  and \emph{information}. We  generate freely the real network by choosing one from regular, random or scale-free. Then we generate a \emph{virtual} network also chosen from the  three benchmark networks, with same average connectivity $\langle k \rangle=2m$.  In order to construct the information  network we overlap the real and the virtual ones and then, for each node, we prune its ``real'' links with probability $q$ and its ``virtual'' ones with probability $1-q$. The overlap between the real and the information networks is thus $1-q$. The information network is not symmetric. 

This procedure allow us to study the effects of the difference between the real network, where the epidemic spreading takes place, and the information one, where actors become aware of the disease, \textit{i.e.}, over which they evaluate the perception of the risk of being infected. An example of multiplex is reported in Fig.~\ref{fig:multiplex}.

\begin{figure}[t!]
\centering
{\includegraphics[width=1\columnwidth]{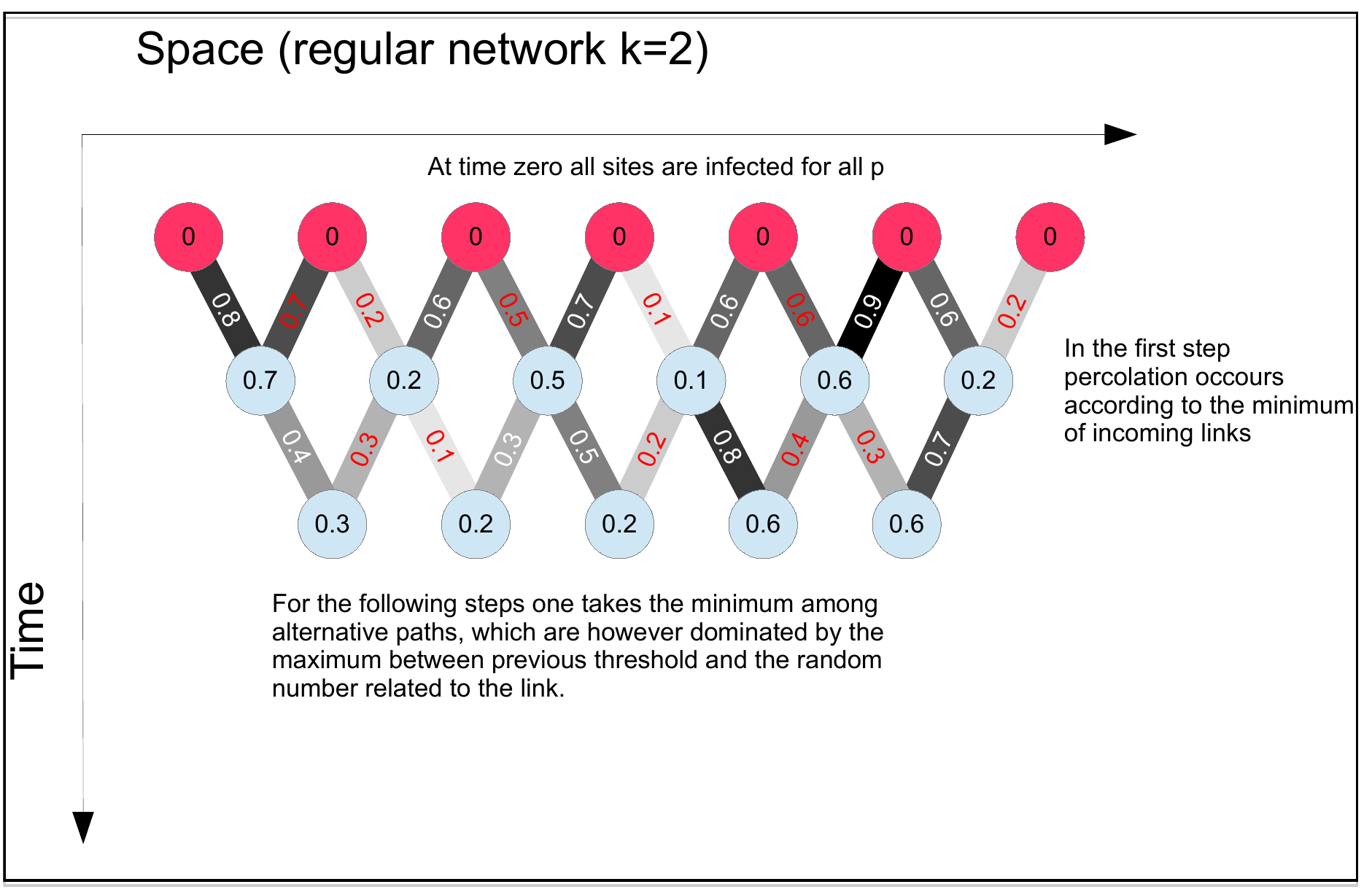}}
\caption{\label{fig:perc_schema} Evolution of the local minimum value of the percolation parameter $p_i$ for a 1D regular network with $k =2$.}
\end{figure}

\section{Infection model and mean-field approximation}
\label{meanfield}
Following Ref.~\cite{riskperception}, we assume that the probability that a site $i$ is infected by a neighbour $j$ is given by 
\[
  u(s, k) = \tau \exp\left(-J\frac{s}{k_i}\right), 
\]
where $\tau$ is the ``bare'' infection probability and $s$ is the number of infected neighbours. The idea is that the perception of the risk, given by the percentage of infected neighbours and modulated by the factor $J$, effectively lowers the infection probability (for instance because people takes more precautions). In the case of information networks, the perception is computed on the mixed real-virtual neighbourhood, while the actual infection process takes places on the real network.

It is possible to derive a simple mean-field approximation for the fixed-$k$ case. Denoting by $c$ the fraction of infected individuals at time $t$ and by $c'$ those at time $t+1$, we have, considering a random network, 
\begin{equation}
\label{mf}
c' = \sum_{s=0}^k \mybinom{k}{s} c ^s (1-c) ^{k-s} p(s, k),
\end{equation}
where $p(s, k)$ is the probability of being infected if there are $s$ out of $k$ infected neighbours. The probability $p$ depends on $u$ as 
\[
 p(s, k) = 1- \bigr[1-u(s,k)\bigl]^s, 
\]
since the infection processes are  independent, although the infection probabilities are coupled by the ``perception''-dependent infection probability $q$, Eq.~\eqref{prp}. 

Near the threshold, the probability $u$ is small, and therefore we can approximate 
\[
 p(s, k) \simeq s u(s,k) = s\tau \exp\left(-J \frac{s}{k}\right).
\]
Replacing $p$ into Eq.~\eqref{mf}, we get
\[
c' = \sum_{s=0}^k \mybinom{k}{s} c^s (1-c)^{k-s} s \tau \exp\left(-J\frac{s}{k}\right),
\]
and setting $a=\exp\left(-Js/k\right)$, 
\[
c' = \tau \sum_{s=0}^k \mybinom{k}{s} c^s (1-c)^{k-s} s a^s,
\]
which gives
\[
c' = \tau a k (c a + 1 - c)^{k-1}.
\]

\begin{figure}[t!]
  \begin{center}
      \includegraphics[width=1\columnwidth]{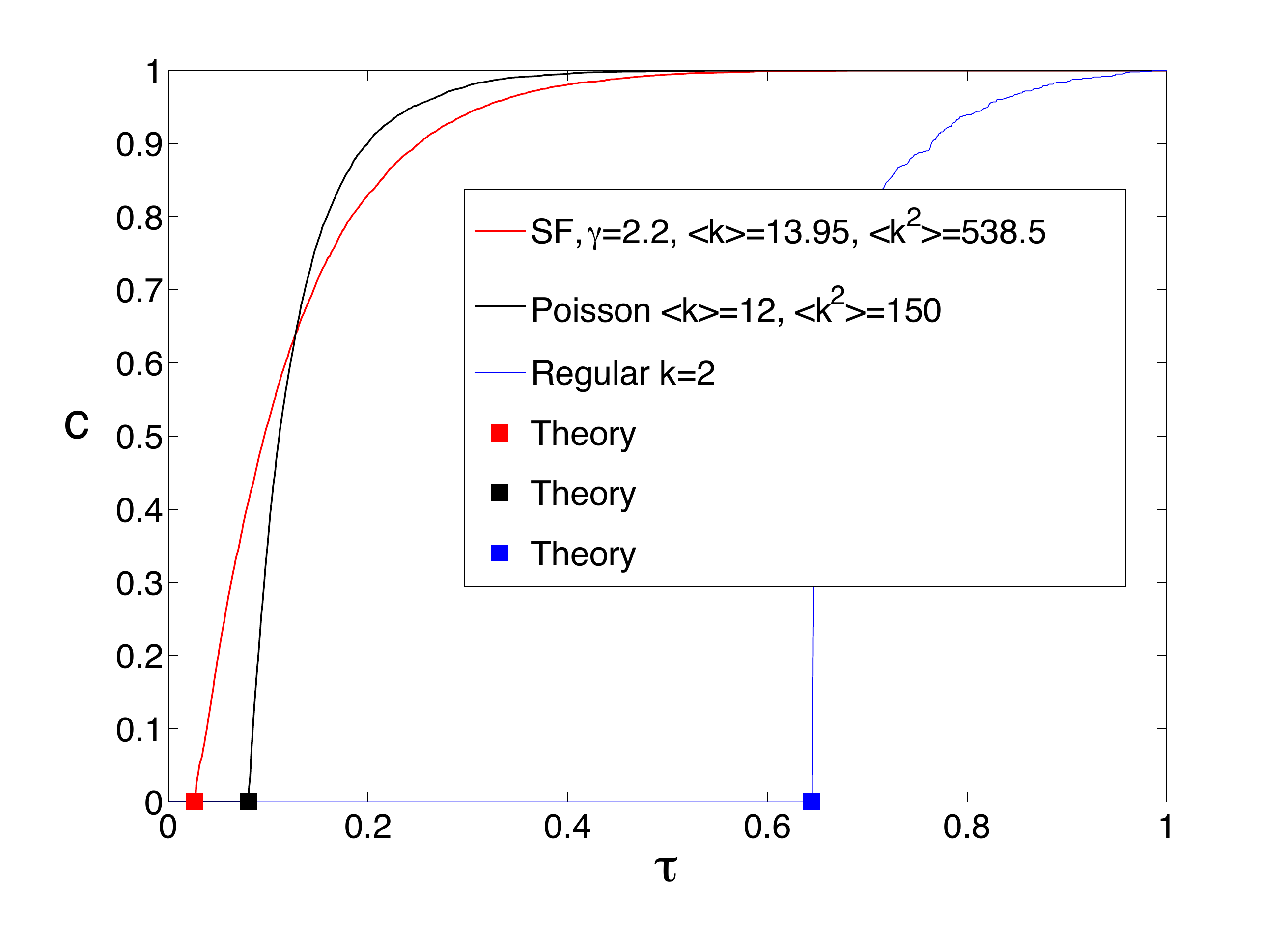}
  \end{center}
    \caption{\label{fig:perc} Asymptotic  number of infected individuals $c$ versus the bare infection probability $\tau$ for the SIS dynamics for different networks, $N=10000$.} 
\end{figure}

The critical threshold $J_c$ corresponds to the stationary state $c'=c$ in the limit $c\rightarrow 0$,\textit{ i.e.}, 
\begin{equation}\label{Jc}
\tau = \frac{1}{k }\exp\left(\frac{J_c}{k}\right) ; \qquad J_c = k \ln(k \tau).
\end{equation}

\begin{figure}
\centerline{\includegraphics[width=1\columnwidth]{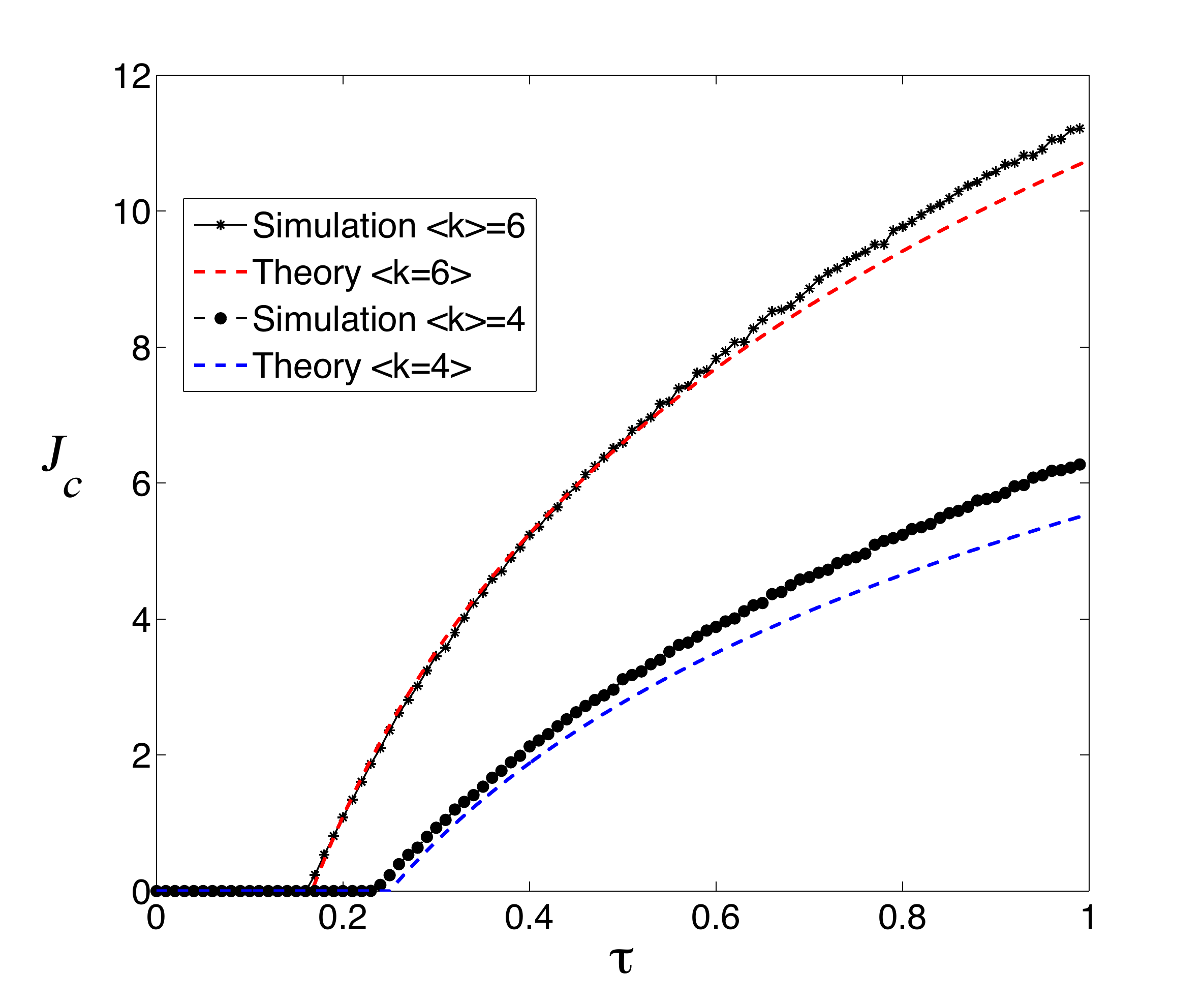}}
\caption{\label{fig:rn-mf-sim}Comparison between mean field approximation and simulations for random networks with different values of $\langle k\rangle$.} 
\end{figure}

This prediction is quite accurate: in Fig.~\ref{fig:rn-mf-sim} the comparison between Eq.~\eqref{Jc} and actual simulations in reported for different values of $\langle k\rangle$ using random networks.

The analysis can be extended to non-homogeneous networks with connectivity distribution $P(k)$ like the  scale-free ones. We can start analysing a node with connectivity $k$
\[
c'_k = \sum_{s_1, s_2, \dots, s_k = 0}^1 \sum_{j_1, j_2,\dots,j_k=0}^\infty \prod_{i=1}^k C(j_i,k)I(s_i, j_i)T(k|s_i),
\] 
where we denote with $i=1,\dots, k$ the neighbours, $s_i=0,1$ is their state (healthy, infected) and $j_i$ their connectivity. $C(j, k)$ is the probability that a node with connectivity $j$ is attached to node with connectivity $k$, $I(s_i, j_i)$ is the probability that the neighbour $i$ is infected, and  $T(k|s_i)$ is the probability that it transmits the infection to the node under investigation.

Clearly, $\sum_j C(j,k) =1$. We use symmetric networks, so $j C(j,k)P(j)=k C(k,j) P_k$ (detailed balance).  For non-assortative networks, $C(j,k)$ does not depend on $k$, and summing over the detailed balance condition, $C(j,k) = j P(j)/\langle k\rangle$.  The quantity $I(s_i, j_i)$ is simply $c_{j_i}^{s_i}(1-c_{j_i})^{1-s_i}$ and $T(k|s_i) = 1-(1-\tau \exp(-J s/k))^s$, where $s=\sum_i s_i$ (risk perception). Near the extinction, $\tau \exp(-J s/k)$ is small and we can approximate $T(k|s_i) = s\tau \exp(-J s/k)$.

Summing up, we have 
\begin{widetext}
\[
\begin{split}
c'_k &= \sum_{s_1, s_2, \dots, s_k = 0}^1 \sum_{j_1, j_2,\dots,j_k=0}^\infty \prod_{i=1}^k \frac{ j_i P(j_i)}{\langle k\rangle}c_{j_i}^{s_i}(1-c_{j_i})^{1-s_i}s\tau\exp\left(-J \frac{s}{k}\right), \\
   &= \sum_{s_1, s_2, \dots, s_k = 0}^1 s\tau \exp\left(-J\frac{ s}{k}\right) \prod_{i=1}^k \sum_{j_i=0}^\infty \frac{ j_i P(j_i)}{\langle k\rangle}c_{j_i}^{s_i}(1-c_{j_i})^{1-s_i}.
 \end{split}
\]
\end{widetext}
Since $s_i=0,1$, the quantity $c_{j_i}^{s_i}(1-c_{j_i})^{1-s_i}$ is either $s_i$ or $1-s_i$. Let us define $\tilde c = \sum_j j c_j/\langle k \rangle$, and we get
\[
\begin{split}
c'_k &= \sum_{s_1, s_2, \dots, s_k = 0}^1 s\tau \exp\left(-J\frac{ s}{k}\right) \prod_{i=1}^k \tilde c^{s_i}(1-\tilde c)^{1-s_i}, \\
 &= \sum_{s=0}^k \binom{k}{s} s\tau \exp\left(-J\frac{ s}{k}\right)  \tilde c^{s}(1-\tilde c)^{k-s},
\end{split}
\]
\textit{i.e.}, 
\[ 
  c'_k= \tilde c k \tau \exp\left(\frac{-J s}{k}\right) \left(\tilde c \exp\left(-\frac{J}{k}\right)  + 1 -\tilde c \right).
\]

\begin{figure}
\centerline{\includegraphics[width=1\columnwidth]{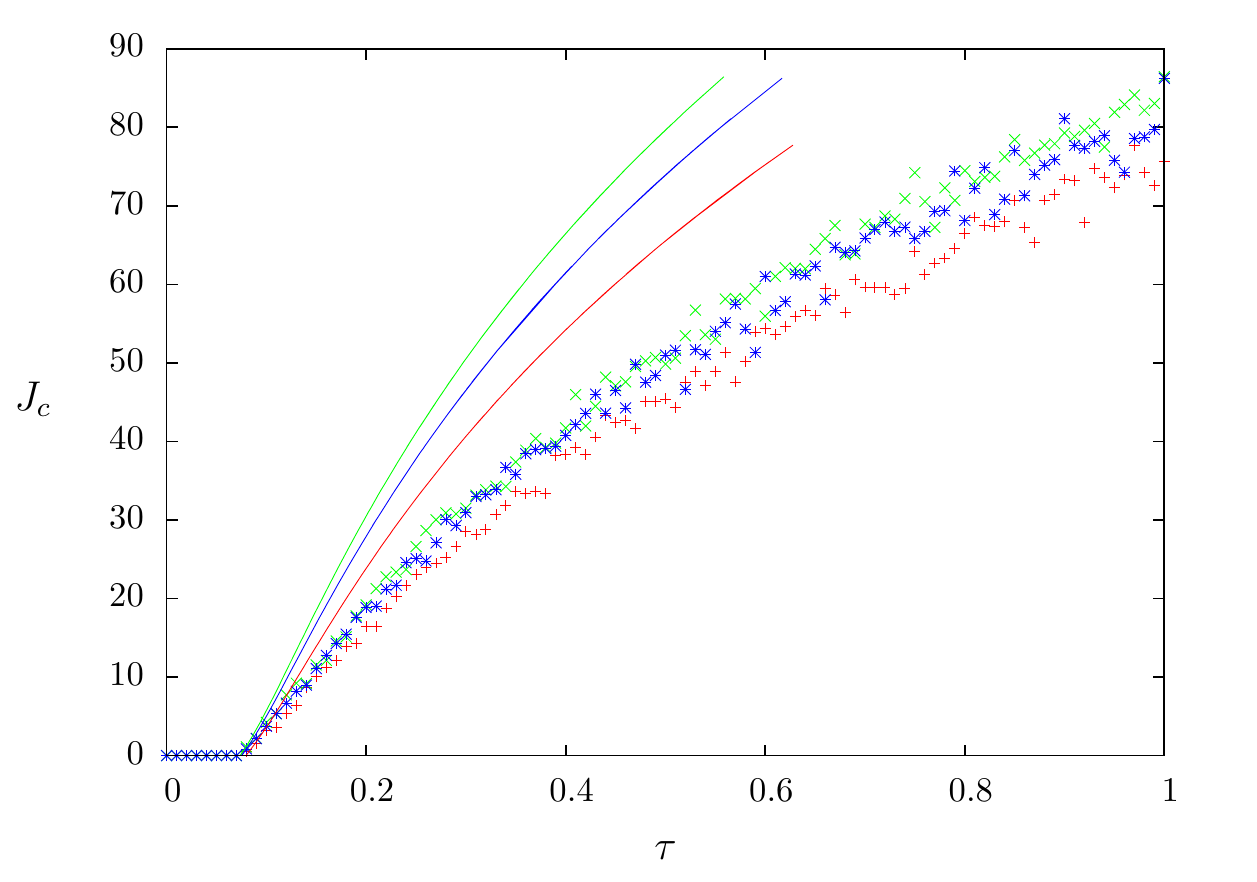}}
\caption{\label{fig:sf-mf-sim}Comparison between the results of the mean field approximation and simulations for three instances of scale-free networks with $\gamma = 2.4$, $N=10000$, $m=2$ and $K=300$. Although the simulations give similar results, the mean-field computations (that coincide with the simulation only for $J_c =0$) are very dependent on the details of the network. }. 
\end{figure}

Near the epidemic threshold $\tilde c \rightarrow 0$, and 
\[
\tilde c' = \frac{1}{\langle k \rangle} \sum_k k c'_k P(k) = \frac{\tau \tilde c}{\langle k \rangle} \sum_k k^2 P(k) \exp\left(-\frac{J}{k}\right). 
\]
The correspondence between $\tau_c$ and $J_c$ is therefore
\begin{equation}\label{eq:tauth}
\tau_c(J_c) = \frac{\langle k \rangle}{ \sum_k k^2 P(k) \exp\left(-\frac{J_c}{k}\right)},
\end{equation}
which, for $J_c=0$, gives the usual relationship $\tau_c = \langle k \rangle/ \langle k^2 \rangle$, that for a sharply peaked $P(k)$ corresponds to Eq.~\eqref{Jc}.

By using a continuous approximation, it is possible to make explicit  the relationship between $\tau$ and $J_c$ in the scale-free case. Eq.~\eqref{eq:tauth} becomes
\begin{equation}\label{eq:tauthc}
\tau_c(J_c) = \frac{\langle k \rangle}{ \int_m^K k^2 P(k) \exp\left(-\frac{J_c}{k}\right)\,\mathrm{d}k},
\end{equation}

Substituting, for the scale free case, $P(k) = A k^{-\gamma}$, where $A$ is the normalization constant so that $\int_m^K P(k) = 1$ 
\[
A = \frac{\gamma-1}{m^{1-\gamma} - K^{1-\gamma}}\simeq (\gamma-1)m^{\gamma-1},
\]
if $K \gg m$ (and $\gamma < 3$). We get 
\[
\langle k \rangle = \frac{\gamma-1}{\gamma-2}\cdot \frac{m^{2-\gamma} - K^{2-\gamma}}{m^{1-\gamma} - K^{1-\gamma}} \simeq \frac{\gamma -1}{\gamma-2} m
\]
for $K\gg m$, and 
\begin{equation}\label{tau}
\tau_c(J_c) =   J_c^{3-\gamma} \left[\Gamma\left(\gamma-3, \frac{J_c}{K}\right) - \Gamma\left(\gamma-3, \frac{J_c}{m}\right)\right],
\end{equation}
where $\Gamma(a,x)$ is the incomplete gamma function. Eq.~\eqref{tau}  diverges for $K\rightarrow\infty$ and thus for infinite networks $J_c =0$ $\forall \tau$. However, real networks always have a cut-off (at least due to the finite number of nodes)~\cite{EvolutionNetworks}. For $J_c=0$ we recover the standard threshold  
\begin{equation}\label{eq:tau0}
\tau_c(0) = \frac{\gamma-3}{\gamma-2}\cdot \frac{m^{2-\gamma} - K^{2-\gamma}}{m^{3-\gamma} - K^{3-\gamma}} \simeq \frac{3-\gamma }{\gamma-2}\cdot \frac{m^{2-\gamma}}{K^{3-\gamma}}.
\end{equation}

The problem of epidemic threshold in finite-size scale-free networks was studied in  Ref.~\cite{FiniteSizeEpidemics}. The conclusions there is that even in finite-size networks the epidemics is hard to stop. Indeed, we find numerically that the epidemics always stops in finite scale-free networks although the required critical value of $J_c$ may be quite large.  

In Fig.~\ref{fig:sf-mf-sim} the comparison between the mean-field prediction, Eq.~\eqref{eq:tauth}, and actual simulations is shown, for three instances of a scale-free network generated with the same parameters. The theoretical prediction coincides with the simulations only for $J_c=0$. Moreover, although the simulation results seems to be sufficiently independent on the details of the generated networks, the theoretical prediction is quite sensible to them. The continuous approximation, Eq.~\eqref{eq:tau0} gives, for $m=2$, $K=300$ and $\gamma =2.4$, a value $\tau(0) \simeq 0.037$, quite different from the computed one $\tau(0) \simeq 0.08$.

\section{The self-organized percolation method}
 Here we show a self-organized percolation method that allows to obtain the critical value of the percolation parameter in a single run, for a given network.  We consider a parallel SIS process, which is equivalent to a directed percolation problem where the directed direction is time. 

Let us denote by $x_i(t)=0,1$  ($0=$healthy, $1=$infected), the percolating variable and by  $p$ the control parameter (percolation probability). 

\subsection{Simple infection (direct percolation)}
Considering the infection probability $\tau$ is fixed, the stochastic evolution process for the network is  defined as 
\begin{equation}
x_i(t+1) = \bigvee_{j=j^{(i)}_1, \dots, j^{(i)}_{ k_i}} [\tau> r_{ij}(t)] x_j(t)  
\label{eq1}
\end{equation}
where $\bigvee$ represents the OR operator and the multiplication represents the AND. The square bracket represents  the truth function, $[\cdot]=1$ if ``$\cdot$'' is true, and zero otherwise. The quantity $r_{ij}(t)$ is a random number between 0 and 1 that varies with $i$, $j$ and $t$.  We want to derive an equation for $\tau_i(t)$, which is the minimum value of $\tau$ for which $x_i(t)$ is infected.  We  can replace $x_i(t)$ by $[\tau> \tau_i(t)]$. Eq.~\eqref{eq1} becomes:
\begin{equation}
\label{eq2}
  [\tau> \tau_i(t+1)] = \bigvee _{j=j^{(i)}_1, \dots, j^{(i)}_{ k_i}} [\tau>r_{ij}(t) ] [\tau>\tau_j(t)].
\end{equation}
Now $[\tau>a ] [\tau>b]$ is equal to $[\tau>\max(a,b)]$ and $[\tau>a] \vee [\tau>b]$ is equal to $[\tau>\min(a,b)]$, therefore Eq.~\eqref{eq2} becomes:
\begin{equation}
\label{eq3}
  [\tau>\tau_i(t+1)] =\left[\tau> \left(\MAX_{j=j^{(i)}_1, \dots, j^{(i)}_{ k_i}} \max\bigl(r_{ij}(t),\tau_j(t)\bigr)\right)\right],
\end{equation}
and therefore we get the equations for the $\tau_i$'s
\begin{equation}
\label{p}
   \tau_i(t+1) = \MIN_{j=j^{(i)}_1, \dots, j^{(i)}_{ k_i}} \max\bigr(r_{ij}(t),\tau_j(t)\bigl).
\end{equation}
Let assume that at time $t=0$ all sites are infected, so that $x_i(0)=1\; \forall \tau$. We can alternatively write $\tau_i(0)=0$ (since the minimum value of $\tau$ for which $x_i(0)=1$ is one for sure. We can therefore iterate Eq.~\eqref{p} and get the asymptotic distribution of $\tau_i$. The minimum  of this distribution gives the critical  value $\tau_c$ for which there is at least one percolating cluster with at least one ``infected'' site  at large times. As usual, $t$ cannot be infinitely large for finite $N$ otherwise there will be surely a fluctuation that will bring the system into the absorbing (healthy $x_i=0$) configuration. A schematic representation of this \textit{modus operandi} is illustrated in  Fig.~\ref{fig:perc_schema}.

\subsection{Infection with risk perception}
Now, let us apply the method to a more difficult problem, for which the percolation probability depends on the fraction of infected sites in the neighbourhood (risk perception). As above, we define the infection probability $u$ as
\begin{equation}
\label{prp}
u(s,k) = \tau \exp\left(-J\cdot \frac{s}{k}\right) 
\end{equation}
 where $\tau$ is the bare infection probability, $s$ is the number of infected neighbors and $k$ is the node connectivity. In this case we want to find the minimum value of the parameter $J$ for which there is no spreading of the infection at large times. 
The quantity $[u>r]=[\tau \exp(-J s/k) > r]$ is equivalent to $[J   < - ( k/s) \ln(r/\tau)]$. Therefore Eq.~\eqref{eq2} is replaced by 
\begin{widetext}
\begin{equation}
\label{pprp}
 [J < J_i(t+1)] 
= \SOR_{j=j^{(i)}_1, \dots, j^{(i)}_{ k_i}} \left[ J < - \frac{k_i}{s_i} \ln\left(\frac{r_{ij}(t)}{\tau}\right)\right] [J < J_j(t)] 
\end{equation}
where
\begin{equation}
\label{s}
    s_i\equiv s_i(J) = \sum_{j=j^{(i)}_1, \dots, j^{(i)}_{ k_i}} x_{j} = \sum_{j=j^{(i)}_1, \dots, j^{(i)}_{ k_i}}       [J_j(t)\ge J]. 
\end{equation}
So
\begin{equation} 
 \label{ppprp}
 [J < J_i(t+1)] =
\SOR_{j=j^{(i)}_1, \dots, j^{(i)}_{ k_i}} \left[ J < - \frac{k_i}{s_i(J_j(t))} \ln\left(\frac{r_{ij}(t)}{\tau}\right)\right] [J < J_j(t)] 
 \end{equation}
and therefore 
\begin{equation}
\label{rp}
J_i(t+1) =   \MAX_{j=j^{(i)}_1, \dots, j^{(i)}_{ k_i}} \min\left( - \frac{k_i}{s_i(J_j(t))} \ln\left(\frac{r_{ij}(t)}{\tau}\right), J_j(t)\right).
\end{equation}
\end{widetext}
Analogously to the previous case, the critical value of $J_c$ is obtained by taking the maximum value of the $J_i(t)$ for some large (but finite) value of $t$.

\subsection{The self-organized percolation method for multiplex networks}

We can now turn to the problem of computing the critical value $J_c$ for a fixed value of $\tau$ if the perception is computed on the information network  which is partially different from the ``real'' where  infection spreads. Here the perception of the importance of the infection,  $\overline{s}_i$,  is computed on the neighbours 
$\overline{j}^{(i)}$ on the information network. The perceived number of infected neighbours  depends on how many of them, in the information network, have a value$J_{\overline{j}}$ larger than that computed in the real network, i.e., 

\begin{equation}
\label{vrp}
J_i(t+1) = \MAX_{j=j^{(i)}_1, \dots, j^{(i)}_{ k_i}} \min\left( - \frac{k_i}{\overline{s}_i(J_j)} \ln\left(\frac{r_{ij}(t)}{\tau}\right), J_j(t)\right),
\end{equation}
where 
\begin{equation}
\label{vvrp}
\overline{s}_i(J) = \sum_{\overline{j}=\overline{j}^{(i)}_1, \dots, \overline{j}_{\overline{k}_i}^{(i)}} [J_{\overline{j}}\ge J].
\end{equation}
In other words: for any value of $J$ in the real neighbourhood one computes how many neighbours $\overline{j}$ in the  \textit{information}  network have $J_{\overline{j}} \ge J$. This is the perceived value of the risk.

\begin{figure}[t!]
\centerline{\includegraphics[width=1\columnwidth]{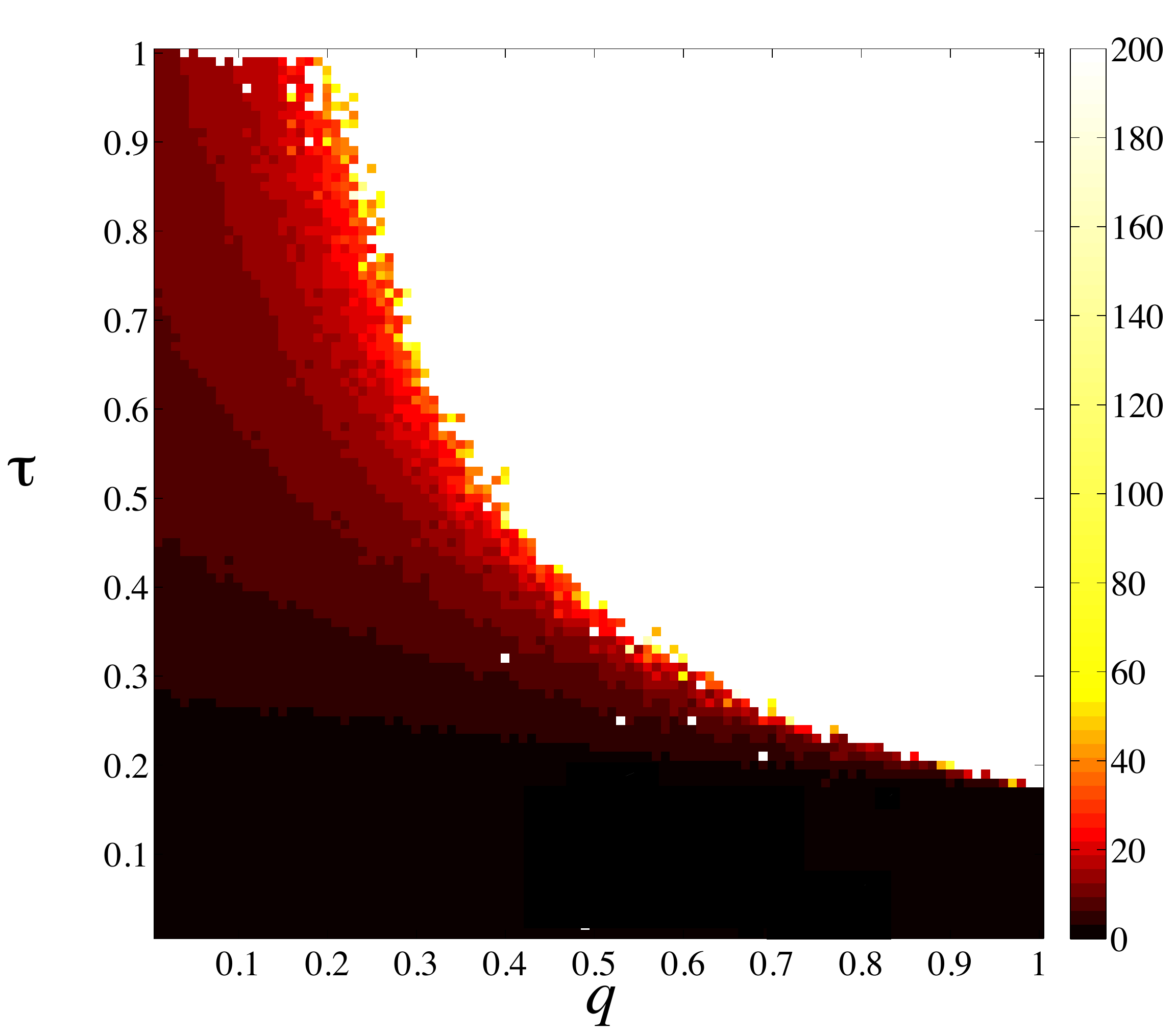}}
    \caption{\label{fig:vrirsk-tau-q} Critical precaution threshold $J_c$ (color code) as a  function of the bare infection $\tau$ and of the  difference between the real and the information network $q$. Here the real and virtual networks are Poissonian (random) with $\langle k\rangle = 6$ and $N=1000$. In the darker region there is always a value of $J_c$ able to stop the epidemics, while in the lighter region the epidemics cannot be stopped. The separation boundary is the stoppability frontier.} 
\end{figure}

\begin{figure}
\centerline{\includegraphics[width=1\columnwidth]{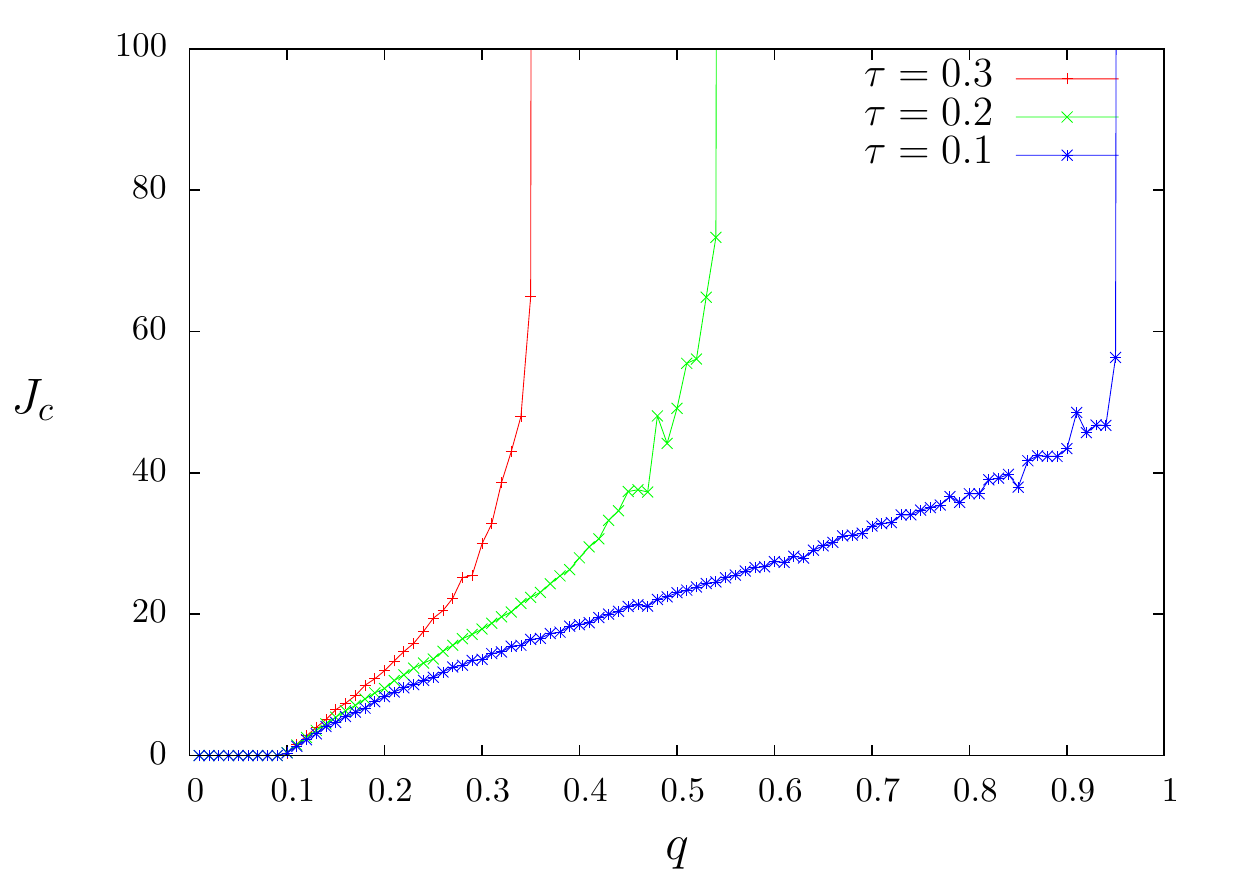}}
\caption{\label{fig:vrisk-rn-sf} Critical precaution threshold $J_c$ versus  the difference  between the real and the information network $q$ for some values of the bare infection $\tau$(from right to left $\tau=0.1, 0.2,0.3$). Random real network and   scale-free virtual network, both with $\langle k\rangle =6$, $N=10000$.} 
\end{figure}

\section{Results}
In this section we show the results of the self-organized percolation method in both single-layered and multiplex networks (with and without risk perception). For our experiments we generally use the network size $N=10000$ and the computational time $T=10000$.

\subsection{Percolation in single-layered networks (SIS dynamics)}
We investigated the  SIS dynamics over regular, Poisson and scale-free networks as shown in Fig.~\ref{fig:perc}. In particular we evaluated the critical epidemic threshold values $\tau_c$ for which there is at least one percolating clusters with at least one infected nodes.

Considering a regular lattice with  connectivity degree $k=2$, we found $\tau_c\simeq 0.6447$ which is compatible with  the results of the bond percolation transition in the Domany-Kinzel model~\cite{Domany-Kinzel}.

In the case of random networks with Poisson degree distributions the critical epidemic threshold $\tau_c= \langle k \rangle /  \langle k^2 \rangle  \simeq \langle k \rangle ^{-1} $ if the distribution is sharp~\cite{epidemic2}. Indeed, for Poisson network with $\langle k \rangle = 12$ the self-organized percolation method gives $\tau_c \simeq 0.08 \simeq 1/12$. 

For a scale-free  network with $\langle k \rangle = 13.95$ and $\langle k^2 \rangle = 538.5$ we get from simulations $\tau_c \simeq 0.026$, in agreement with the expected value.

\subsection{The effects of risk perception in SIS dynamics}
We investigate the effects of risk perception in the previous simple model of epidemic spreading. 
The results are quite interesting if compared with the simple SIS dynamics (Fig.~\ref{fig:perc}) by inserting the risk perception it is possible to stop the epidemic for every value of the bare infection probability $\tau$ up to $\tau=1$.  let us consider for instance the case of random networks with $\langle k\rangle=6$. For the simple infection process we found  a critical value  $\tau_c=0.165$ (Fig.~\ref{fig:perc}). As shown in  Fig.~\ref{fig:rn-mf-sim}, beyond  this value of $\tau_c$ the epidemics can still be stopped if all agents adopt a sufficiently high precaution level $J$. The same consideration can be done also for the other scenarios.

 \begin{figure}[t!]
\centerline{\includegraphics[width=1\columnwidth]{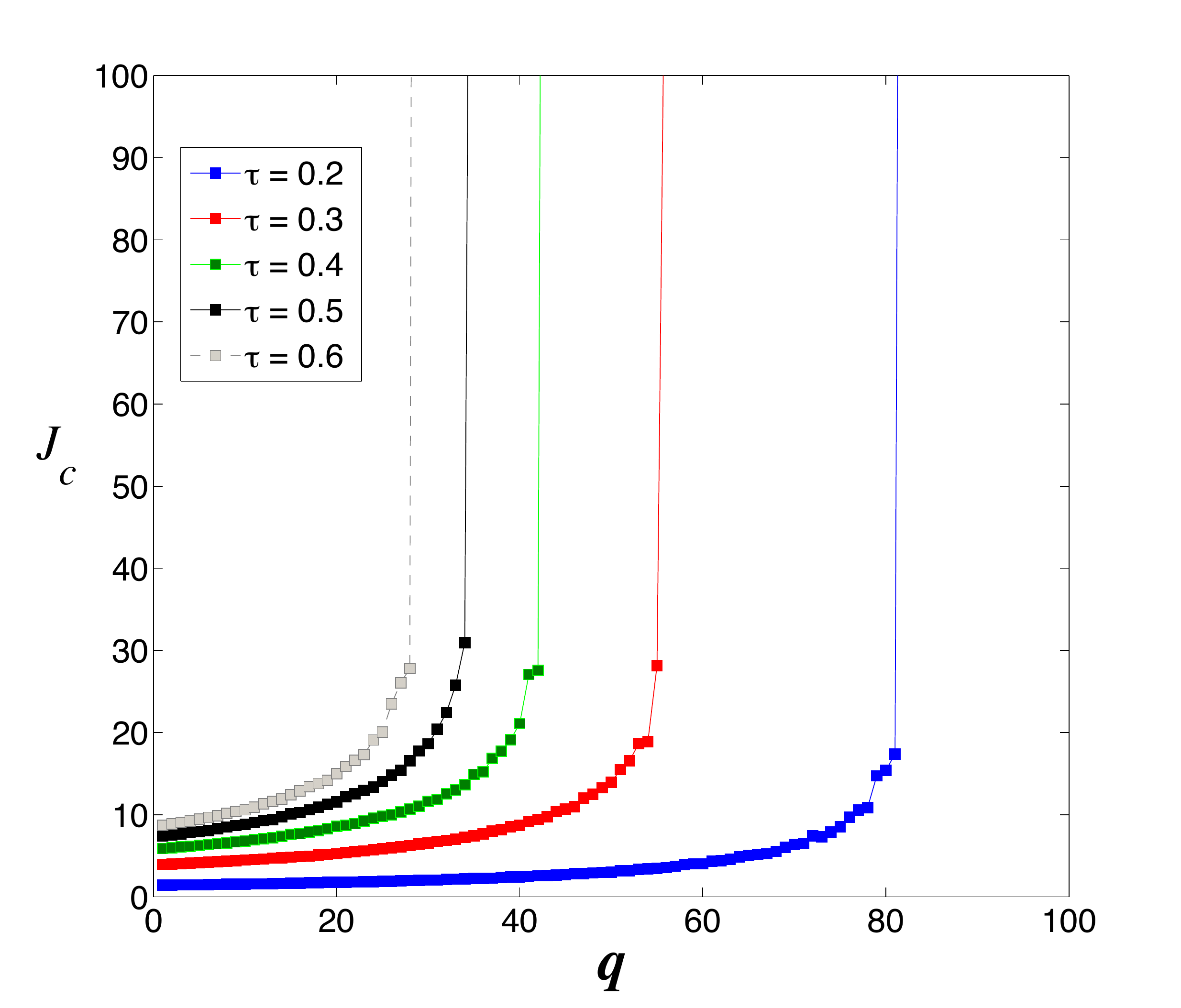}}
    \caption{\label{fig:vrirsk-rn-rn} Critical precaution threshold $J_c$ versus the difference  between the real and the information network $q$ for some values of the  bare infection $\tau$ (from right to left: $\tau = 0.2, 0.3, \dots, 0.6$).  Random real  and  virtual networks, both with $\langle k \rangle =6$ and $N=10000$.} 
\end{figure}

\subsection{Multiplex risk perception}
The phase diagram for the risk-perception over SIS dynamics in multiplex networks is reported in  Fig.~\ref{fig:vrirsk-tau-q}. The general shape of this phase diagram can be understood considering that a given node of real connectivity $k_r$ is connected, in the information network, to  $(1-q)k_r$ real neighbours, the rest being virtual ones. At the threshold, the global fraction of infected sites is small. For the spreading of the epidemics, the important sites are those that have a infected real neighbour. It might be assumed that the virtual neighbours, being uncorrelated with the real ones, do not contribute at all to the risk perception. Among the real neighbours, the fraction $qk_r$ replaced by virtual ones has become invisible, and so the perception decreases by   a factor $q$.  For a given value of the perception, the infection is stoppable only if the infectiousness is also  decreased by a factor $q$. Indeed, the shape of the stoppability frontier of Fig.~\ref{fig:vrirsk-tau-q} resembles that of an hyperbola $\tau q = \text{const}$. 

The general trend is that,  increasing the difference $q$ between the information network and the real one it becomes harder to stop an epidemics. It is interesting to investigate this transition. As we can see in Fig.~\ref{fig:vrisk-rn-sf} for a real random and ghost scale-free network, this transition is quite sharp, especially for low values of $\tau$.  A similar scenario holds for a mixture of  real and  ghost random networks, as shown in Fig.~\ref{fig:vrirsk-rn-rn}.

\section{Conclusions}

We  investigated the interplay between epidemic spreading and risk perception on multiplex networks, exploiting mean-field approximations and a self-organized method, that automatically gives the percolation threshold in just one simulation. 
We focused on multiplex networks, considering that  people get infected by contacts in real life but often gather information  from an information networks, that may be quite different from the real ones. The main conclusion is that the similarity between the real and information networks determine the possibility of stopping the infection for a sufficiently high precaution level: if the networks are too different there is no mean of avoiding the epidemics. Moreover, especially for low values of the bare infection probability, this transition occurs sharply without evident forerunners. This last observation remarks that, although the virtual world has indeed the advantage of allowing a fast diffusion of information, real epidemics still propagate in the real world. This is of particular importance for ``neglected'' diseases, possibly diffused in marginalized parts of the population that have little access to Internet, and that in any case are not part of the ``neighbourhood'' of ``real neighbours'' belonging to other social classes.

\section*{Acknowledgements}
FB acknowledges partial financial support from the EU projects 288021 (EINS -- Network of Excellence in Internet Science) and  611299 (SciCafe2.0). The author are grateful to  Dr. Nicola Perra for helpful discussions.

\bibliography{biblio}{}

\begin{thebibliography}{10}

\bibitem{storia0}
\href{http://www.health.com/health/gallery/0,,20307381_2,00.html}{Pandemic
  Scares Throughout History}.
\newblock {\em Health Magazine}, 2013.

\bibitem{pandemic}
Wikipedia.
\newblock \url{http://en.wikipedia.org/wiki/Pandemic}, 2013.

\bibitem{storia}
\href{http://www.dailymail.co.uk/news/article-1242147/The-false-pandemic-Drug-firms-cashed-scare-swine-flu-claims-Euro-health-chief.html}{The
  ``false'' pandemic: Drug firms cashed in on scare over swine flu, claims Euro
  health chief}.
\newblock {\em Daily Mail}, 2010.

\bibitem{epidemic1}
C.~Moore and M.~E.~J. Newman.
\newblock Epidemics and percolation in small-world networks.
\newblock {\em Phys. Rev. E}, 61:5678--5682, 2000.

\bibitem{epidemic2}
R.~Pastor-Satorras and A.~Vespignani.
\newblock Epidemic spreading in scale-free networks.
\newblock {\em Phys. Rev. Lett.}, 86:3200--3203, 2001.

\bibitem{epidemic3}
M.~E.~J. Newman.
\newblock Exact solutions of epidemic models on networks.
\newblock Working Papers 01-12-073, Santa Fe Institute, December 2001.

\bibitem{epidemic4}
R.~M. May and A.~L. Lloyd.
\newblock Infection dynamics on scale-free networks.
\newblock {\em Phys. Rev. E}, 64:066112, 2001.

\bibitem{epidemic5}
R.~Pastor-Satorras and A.~Vespignani.
\newblock Immunization of complex networks.
\newblock {\em Phys. Rev. E}, 65:036104, 2002.

\bibitem{wikipedia}
Wikipedia.
\newblock \url{http://en.wikipedia.org/wiki/Lazaretto}, 2013.

\bibitem{lazzaretti}
R.~J. Palmer.
\newblock {\em L'azione della repubblica di Venezia nel controllo della peste.
  Lo sviluppo della politica governativa, Venezia e la peste 1348–1797}.
\newblock Marsilio Editori, Venice (Italy), 1979.

\bibitem{riskperception}
F~Bagnoli, P.~Li\`o, and L.~Sguanci.
\newblock Risk perception in epidemic modeling.
\newblock {\em Phys. Rev. E}, 76:061904, 2007.

\bibitem{virtual_inf}
J.~Ginsberg, M.~Mohebbi, R.~Patel, L.~Brammer, M.~Smolinski, and L.~Brilliant.
\newblock Detecting influenza epidemics using search engine query data.
\newblock {\em Nature}, 457:1012--1014, 2009.

\bibitem{virtual_inf1}
D.~Scanfeld, V.~Scanfeld, and E.~L. Larson.
\newblock Dissemination of health information through social networks: Twitter
  and antibiotics.
\newblock {\em American Journal of Infection Control}, 38(3):182 -- 188, 2010.

\bibitem{virtual_inf2}
C.~Chew and G.~Eysenbach.
\newblock Pandemics in the age of twitter: Content analysis of tweets during
  the 2009 h1n1 outbreak.
\newblock {\em PLoS ONE}, 5(11):e14118, 11 2010.

\bibitem{article}
\href{http://stateofthemedia.org}{The State of the News Media}.
\newblock {\em The Pew Research Center's project for Excellence in Journalism},
  2010.

\bibitem{multiplex1}
M.~Kurant and P.~Thiran.
\newblock Layered complex networks.
\newblock {\em Phys. Rev. Lett.}, 96:138701, 2006.

\bibitem{multiplex2}
P.~J. Mucha, T.~Richardson, K.~Macon, M.A. Porter, and J.-P. Onnela.
\newblock Community structure in time-dependent, multiscale, and multiplex
  networks.
\newblock {\em Science}, 328(5980):876--878, 2010.

\bibitem{multiplex3}
M.~Szell, R.~Lambiotte, and S.~Thurner.
\newblock {Multirelational Organization of Large-scale Social Networks in an
  Online World}.
\newblock 2010.

\bibitem{multiplex4}
A.~Arenas S.~Lozano, X-P.~Rodriguez.
\newblock {Evolution of cooperation in multiplex networks.}
\newblock {\em Scientific reports}, 2, 2012.

\bibitem{multiplex5}
G.~Bianconi.
\newblock Statistical mechanics of multiplex networks: Entropy and overlap.
\newblock {\em Phys. Rev. E}, 87:062806, 2013.

\bibitem{interdependent1}
S.~V. Buldyrev, R.~Parshani, G.~Paul, H.~E. Stanley, and S.~Havlin.
\newblock {Catastrophic cascade of failures in interdependent networks}.
\newblock {\em Nature}, 464(7291):1025--1028, 2010.

\bibitem{interdependent2}
J.~Gao, S.~V. Buldyrev, H.~E. Stanley, and S.~Havlin.
\newblock {Networks formed from interdependent networks}.
\newblock {\em Nat Phys}, 8(1):40--48, 2012.

\bibitem{multiplex9}
C.~Granell, S.~G\'omez, and A.~Arenas.
\newblock Dynamical interplay between awareness and epidemic spreading in
  multiplex networks.
\newblock {\em Phys. Rev. Lett.}, 111:128701, Sep 2013.

\bibitem{multiplexnuovo}
C.~{Granell}, S.~{G\'omez}, and A.~{Arenas}.
\newblock {Competing spreading processes on multiplex networks: awareness and
  epidemics}.
\newblock {\em ArXiv e-prints 1405.4480}, May 2014.

\bibitem{bagnoli_rech}
F.~Bagnoli, P.~Palmerini, and R.~Rechtman.
\newblock Algorithmic mapping from criticality to self-organized criticality.
\newblock {\em Phys. Rev. E}, 55:3970--3976, Apr 1997.

\bibitem{EvolutionNetworks}
S.~N. Dorogovtsev and J.F.F. Mendes.
\newblock Evolution of networks.
\newblock {\em Advances in Physics}, 51:1079--1187, 2002.

\bibitem{FiniteSizeEpidemics}
R.~Pastor-Satorras and A.~Vespignani.
\newblock Epidemic dynamics in finite size scale-free networks.
\newblock {\em Physical Review E}, 65(035108), 2002.

\bibitem{Domany-Kinzel}
E.~Domany and W.~Kinzel.
\newblock Equivalence of cellular automata to ising models and directed
  percolation.
\newblock {\em Phys. Rev. Lett.}, 53:311--314, 1984.

\end{thebibliography}
\bibliographystyle{unsrt}

\end{document}